\documentclass[preprint,12pt]{elsarticle}
\biboptions{sort&compress}

\usepackage{amssymb}
\usepackage{amsmath}
\usepackage{bm}
\usepackage{physics}
\usepackage{etoolbox}
\usepackage{xcolor}
\usepackage[normalem]{ulem}
\newcommand{\red}[1]{\textcolor{red}{#1}}

\usepackage{hyperref}
\usepackage{doi}

\journal{Journal of Magnetism and Magnetic Materials}

\begin{document}

\begin{frontmatter}

\title{Spin Seebeck effect in magnetic junctions with a compensated ferrimagnet}
\author[inst1]{Xin Theng Lee} 
\author[inst1]{Takahiro Misawa}
\author[inst2,inst3,inst4,inst5]{Mamoru Matsuo}
\author[inst1]{Takeo Kato} 

\affiliation[inst1]{organization={Institute for Solid State Physics, The University of Tokyo},
addressline={5-1-5 Kashiwanoha}, 
city={Kashiwa},
postcode={277-8581}, 
state={Chiba},
country={Japan}}
\affiliation[inst2]{organization={Advanced Science Research Center, Japan Atomic Energy Agency},
addressline={2-4 Shirakata-Shirane, Tokai-mura}, 
city={Naka-gun},
postcode={319-1195}, 
state={Ibaraki},
country={Japan}}
\affiliation[inst3]{organization={Kavli Institute for Theoretical Sciences, University of Chinese Academy of Sciences},
addressline={No. 19 Yuquan Road, Haidian District}, 
city={Beijing},
postcode={100190}, 
country={China}}
\affiliation[inst4]{organization={CAS Center for Excellence in Topological Quantum Computation, University of Chinese Academy of Sciences},
addressline={No. 80 Zhongguancun East Road, Haidian District}, 
city={Beijing},
postcode={100190}, 
country={China}}
\affiliation[inst5]{organization={RIKEN Center for Emergent Matter Science (CEMS)},
addressline={2-1 Hirosawa}, 
city={Wako},
postcode={351-0198},
state={Saitama},
country={Japan}}

\begin{abstract}
Compensated ferrimagnets enable ferromagnet-like spin transport without net magnetization.
We study the spin Seebeck effect in a 
compensated ferrimagnet/normal-metal junction using a four-sublattice model in which sublattice inequivalence arises from differences in exchange couplings, in contrast to the previously studied anisotropy-based mechanism. 
Within the nonequilibrium Green's function framework, we show that isotropic magnon splitting generates a robust spin current with a magnitude comparable to that in standard ferromagnetic junctions.
We also demonstrate that the spin Seebeck effect vanishes 
in altermagnet junctions under identical conditions, thereby establishing compensated ferrimagnets as uniquely suited for thermal spin-current generation among magnetically compensated systems.
These results provide a theoretical basis for the applications of compensated ferrimagnets with exchange-coupling asymmetry as stray-field-free spin-current sources in spintronic devices.
\end{abstract}


\end{frontmatter}

\section{Introduction}

Recent studies on collinear antiferromagnets with sublattice-dependent environments have revised the conventional perspective that zero net magnetization necessarily implies spin-degenerate electronic and magnon bands.
However, even in such collinear antiferromagnetic systems, spin splitting can still emerge when spin-up and spin-down sites become inequivalent
~\cite{Leuken_PRL1995,Akai_PRL2006,Kawamura_PRL2024,Liu_PRL2025,Smejkal2022,Smejkal_PRX2022b,Mazin2022,Noda_PCCS2016,Ahn_PRB2019,Hayami2020,Yuan2023,Naka2019}.
Altermagnets and compensated ferrimagnets are representative classes of such systems, but their spin-splitting structures are qualitatively different.
Altermagnets exhibit anisotropic momentum-dependent splitting determined by the crystallographic relationship between the sublattices~\cite{Smejkal2022,Smejkal_PRX2022b,Mazin2022,Noda_PCCS2016,Naka2019,Ahn_PRB2019,Hayami2020,Yuan2023},
whereas compensated ferrimagnets exhibit nearly isotropic ($s$-wave-like) splitting due to complete sublattice inequivalence~\cite{Leuken_PRL1995,Akai_PRL2006,Kawamura_PRL2024,Liu_PRL2025}.

Both classes are promising for spintronics as they combine vanishing stray fields with finite spin splitting.
While bulk transport in altermagnets has been extensively studied~\cite{Naka2019,Hodt2024,Cui2023}, including thermally driven spin transport~\cite{Uchida2008,Jaworski2010,Slachter2010,Uchida2011,Jaworski2012,Adachi2013}, analyses of compensated ferrimagnets remain limited.
Since practical spintronic functionalities are realized at heterointerfaces, clarifying spin transport in compensated ferrimagnet/metal junctions is essential.
This motivates a focused study of compensated ferrimagnets, where isotropic spin-splitting is expected to enable ferromagnet-like spin-current generation.

Sublattice inequivalence in compensated ferrimagnets can arise from two distinct microscopic mechanisms: (i) unequal single-ion anisotropies and (ii) inequivalent exchange couplings.
In Ref.~\cite{Lee2025}, we studied the spin Seebeck effect and spin pumping in a two-sublattice model based on mechanism~(i).
The simplicity of the two-sublattice description makes the essential physics of spin transport in compensated ferrimagnets transparent.
However, single-ion anisotropies are 
typically small and difficult to tune independently in experiments, particularly in organic materials.
By contrast, exchange-coupling asymmetry [mechanism~(ii)] arises naturally in organic antiferromagnets~\cite{Kawamura_PRL2024} and van der Waals heterostructures~\cite{Liu_PRL2025}, where 
distinct molecular or layer environments of the sublattices lead to inequivalent magnetic interactions.
Moreover, since exchange couplings are typically much larger than single-ion anisotropies, this mechanism can produce larger magnon splitting and potentially larger spin-current signals.

\begin{figure}[tb]
    \centering
    \includegraphics[width=1.0\linewidth]{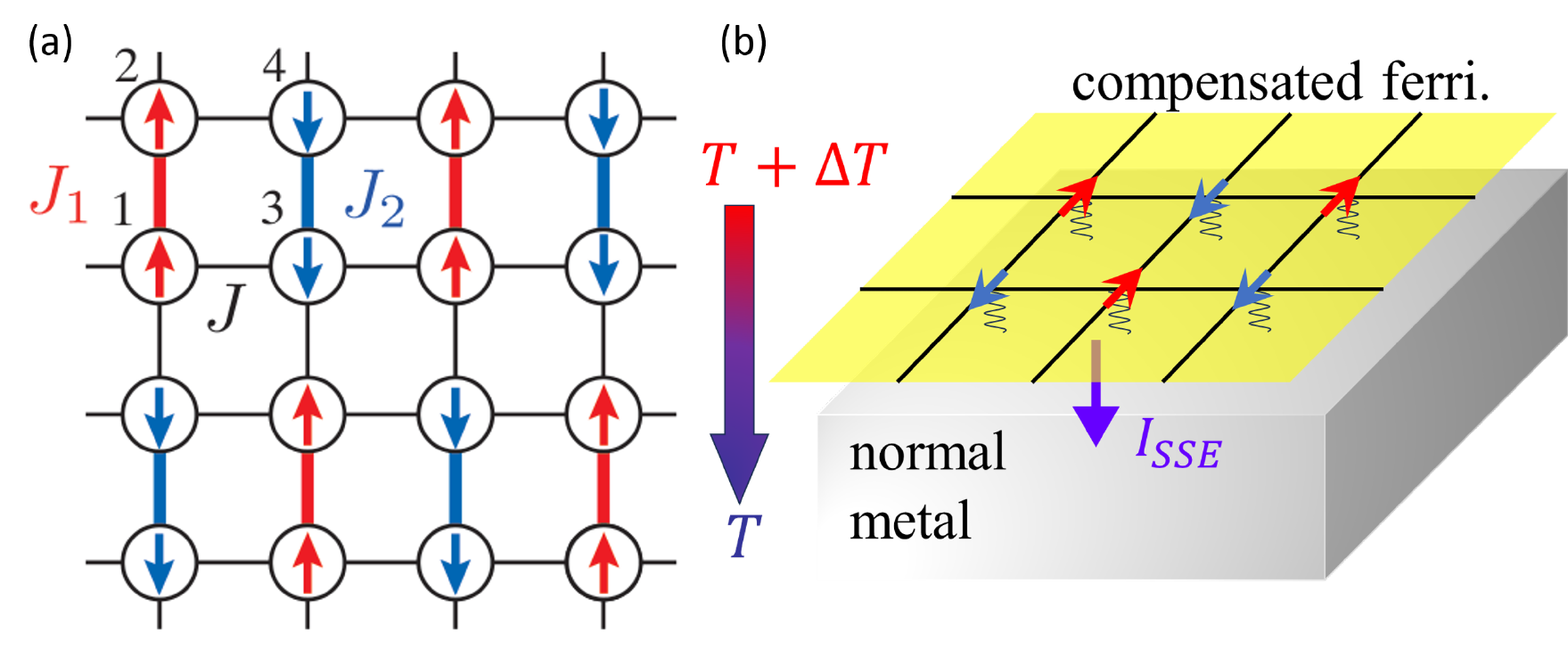}
    \caption{(a) Quantum Heisenberg model for a compensated ferrimagnet. The black lines indicate 
    bonds with 
    antiferromagnetic exchange interaction 
    $J$ ($>0$), while the red and blue lines indicate 
    bonds with 
    ferromagnetic exchange interactions 
    $J_1$ and $J_2$ ($J_1, J_2<0$).
    The sublattice index $\nu$ ($=1,2,3,4$) is defined as shown.
    We consider a magnetic ordering in which 
    spins align in the $+z$ direction for $\nu = 1,2$ and in the $-z$ direction for $\nu = 3,4$. (b) Heterostructure of  compensated ferrimagnet (CF) and  normal metal (NM) with temperature difference $\Delta T$ across the junction.}
    \label{fig:model}
\end{figure}

Although Ref.~\cite{Lee2025} demonstrated that spin pumping in CF/NM junctions exhibits a N\'{e}el-state-dependent resonance-frequency shift when sublattice-dependent single-ion anisotropies are present, this mechanism requires a finite anisotropy gap at the $\Gamma$ point.
In 
organic compounds~\cite{Kawamura_PRL2024} and van der Waals heterostructures~\cite{Liu_PRL2025}, which are the primary targets of the present exchange-coupling-based model, single-ion anisotropies are expected to be small and not independently tunable.
We therefore focus 
on the spin Seebeck effect, which probes magnon splitting throughout the Brillouin zone and does not rely on a finite anisotropy gap.

In this study, we consider the spin Seebeck effect (SSE)~\cite{Uchida2008,Jaworski2010,Slachter2010,Uchida2011,Jaworski2012,Adachi2013} in compensated ferrimagnets (CFs) based on exchange-coupling asymmetry [mechanism~(ii)] with negligible single-ion anisotropies, employing a minimal lattice model as shown in Fig.~\ref{fig:model}(a).
In contrast to the two-sublattice anisotropy-based model of Ref.~\cite{Lee2025}, the present model employs a four-sublattice unit cell on a square lattice with two inequivalent ferromagnetic exchange couplings $J_1$ and $J_2$.
Setting $J_1 \ne J_2$ realizes sublattice inequivalence and captures the key features of CFs in a strong-coupling form~\cite{Kawamura_PRL2024}.
We formulate the spin current induced by the spin Seebeck effect in a CF/normal-metal (NM) junction, as shown in Fig.~\ref{fig:model}(b), using the nonequilibrium Green's function method.
We show that inequivalent sublattices produce isotropic magnon splitting and thereby enable sizable thermal spin currents.
In contrast to the anisotropy-based model, where the spin Seebeck signal is two orders of magnitude smaller than that in ferromagnetic junctions~\cite{Lee2025}, we demonstrate that the exchange-coupling-based model yields a spin-current magnitude comparable to that of standard ferromagnetic junctions.
We also show that the spin Seebeck effect vanishes identically in altermagnet/normal-metal junctions under uniform interfacial coupling, highlighting the distinct advantage of compensated ferrimagnets for thermal spin-current generation.
These results demonstrate that CFs can serve as efficient spin-current sources without net magnetization.

We clarify the terminology used in this paper.
The compensated ferrimagnet (CF) discussed here does {\it not} refer to a ferrimagnet 
at its compensation temperature, where the net magnetic moment 
cancels out due to thermal fluctuations at a specific temperature~\cite{Cramer2017,Gepraegs2016,Li2022}. 
Rather, we focus on 
an intrinsically compensated state realized by inequivalent sublattices with opposite spin alignments, which exhibits zero net magnetization at zero temperature while exhibiting isotropic spin splitting.
Our analysis therefore focuses on the microscopic spin transport properties of such intrinsically compensated systems.

The paper is organized as follows.
In Sec.~\ref{sec:model}, we introduce a theoretical model for the CF/NM junction. We also derive the magnon dispersion relations of the CF. In Sec.~\ref{sec:formulation}, we derive an expression for the spin current induced by the SSE using nonequilibrium Green's functions. 
Sec.~\ref{sec:result} presents numerical evaluations of thermally driven spin currents. Finally, Sec.~\ref{sec:summary} summarizes the main findings and discusses potential implications for spintronic applications. Five appendices provide detailed calculations of magnon dispersions, nonequilibrium Green's functions, and the spin current.

\section{Model}
\label{sec:model}

In this section, we introduce a model for 
the spin Seebeck effect in a CF/NM junction
shown in Fig.~\ref{fig:model}(b).
The total Hamiltonian is given by
\begin{align}
H = H_{\text{CF}} + H_{\text{NM}} + H_{\text{int}},
\end{align}
where $H_{\text{CF}}$, $H_{\text{NM}}$, and $H_{\text{int}}$ represent the compensated ferrimagnet, the normal metal, and the interfacial exchange coupling, respectively.
In the following, we give explicit forms for these terms.

\subsection{Compensated ferrimagnets}

We consider a quantum Heisenberg model on a square lattice with exchange interactions shown in Fig.~\ref{fig:model}(a).
The Hamiltonian is given by
\begin{align}
H_{\text{CF}} &= \sum_{\langle {\bm r},{\bm r}'\rangle} J_{\bm{r},\bm{r}'} \bm{S}_{\bm{r}} \cdot \bm{S}_{\bm{r}^{\prime}} ,\label{rawmodelequation}
\end{align}
where $\bm{S}_{\bm{r}} = (S^x_{\bm r},S^y_{\bm r},S^z_{\bm r})$ is the spin operator of magnitude $S_0$ and $\sum_{\langle {\bm r},{\bm r}'\rangle}$ denotes the sum over bonds connecting 
nearest-neighbor sites at $\bm r$ and $\bm r'$.
The exchange interaction strength $J_{\bm{r},\bm{r}'}$ takes the values $J$, $J_1$, or $J_2$ depending on the bond type as shown in Fig.~\ref{fig:model}(a).  
We assume that $J>0$ and $J_1, J_2 < 0$, leading to the magnetic ordering shown in Fig.~\ref{fig:model}(a) at zero temperature.
This ground state has a unit cell of $2\times 2$ sites and 
each site is specified by a combination of the lattice vector $\bm R$ and the sublattice index $\nu$ ($=1,2,3,4$), as defined in Fig.~\ref{fig:model}(a).
The primitive lattice vectors are given by $\bm a_1 = (2a,0)$ and $\bm a_2 = (a,2a)$, where $a$ is the lattice constant.
The lattice vector is expressed as ${\bm R} = m_1 {\bm a}_1 + m_2 {\bm a}_2$, where $m_1$ and $m_2$ are integers.
When $J_1 \ne J_2$, the two sites with spins aligned along $+z$ are not equivalent to the other two sites with spins aligned along $-z$ under any symmetry operation that combines time reversal with a spatial transformation, thereby realizing a CF insulating state.

To capture the fundamental features of the CF, we employ the spin-wave approximation based on the Holstein--Primakoff transformation:
\begin{align}
S_{\nu \bm{R}}^z &\approx 
\begin{cases}
S_0 - b_{\nu \bm{R}}^{\dagger}b_{\nu \bm{R}} & (\nu = 1,2), \\
-S_0 + b_{\nu \bm{R}}^{\dagger}b_{\nu \bm{R}} & (\nu = 3,4), 
\end{cases} \label{spinwaveapprox1} \\
S_{\nu \bm{R}}^+ & \approx 
\begin{cases}
\sqrt{2S_0}b_{\nu \bm{R}} & (\nu = 1,2) , \\
\sqrt{2S_0}b_{\nu \bm{R}}^\dagger & (\nu = 3,4) , 
\end{cases}
\label{spinwaveapprox2} \\
S_{\nu \bm{R}}^- & = (S_{\nu \bm{R}}^+)^\dagger ,
\label{spinwaveapprox3}
\end{align}
where $S_{\nu {\bm R}}^\pm = S_{\nu {\bm R}}^x \pm i S_{\nu {\bm R}}^y $ and $b_{\nu{\bm R}}^\dagger$ and $b_{\nu{\bm R}}$ are bosonic magnon creation and annihilation operators at the site specified by $\nu$ and ${\bm R}$.
Using this approximation, the Hamiltonian is rewritten in quadratic form in terms of the magnon creation and annihilation operators as
\begin{align}\label{modelequation}
    H_{\text{CF}} = \frac{1}{2} \sum_{\bm{k} \in \Lambda} {\bm \Psi}^{\dagger} M_{\bm{k}} {\bm \Psi},
\end{align}
where ${\bm \Psi} = (b_{1\bm{k}},b_{2\bm{k}},b_{3\bm{k}},b_{4\bm{k}},b_{1,-\bm{k}}^\dagger,b_{2,-\bm{k}}^\dagger,b_{3,-\bm{k}}^\dagger,b_{4,-\bm{k}}^\dagger)^t$, $\Lambda$ denotes the first Brillouin zone, and $b_{\nu {\bm k}}$ is the Fourier transform of $b_{\nu {\bm R}}$.
The explicit form of $M_{\bm k}$ is given in \ref{appex:DerivationofHamiltonian}.

To diagonalize the Hamiltonian, we apply a Bogoliubov transformation by introducing new bosonic operators $a_{\nu {\bm k}}$ defined as~\cite{Shindou2013,Toth2015}
\begin{align}
{\Phi}_j = \sum_{\nu=1}^8 (T_{\bm k}^\dagger)_{j\nu} \Psi_\nu, \quad (j=1,2,\cdots,8).
\end{align}
Here, ${\bm \Phi} =(a_{1\bm{k}},a_{2\bm{k}},a_{3\bm{k}},a_{4\bm{k}},a_{1,-\bm{k}}^\dagger,a_{2,-\bm{k}}^\dagger,a_{3,-\bm{k}}^\dagger,a_{4,-\bm{k}}^\dagger)^t$ and $T_{\bm k}$ is an $8\times 8$ matrix satisfying $T_{\bm k}^\dagger I_- T_{\bm k} = I_-$, where $I_- = \mathrm{diag}(1,1,1,1,-1,-1,-1,-1)$.
The Hamiltonian is then diagonalized as
\begin{align}
    H_{\text{CF}} = \sum_{\bm{k} \in \Lambda} \sum_{j=1}^4 \varepsilon_{j{\bm k}} a^{\dagger}_{j {\bm k}} a_{j {\bm k}} +{\rm const.} ,\label{eq:diagonalized}
\end{align}
where $\varepsilon_{j\bm k}$ denotes the energy dispersion of the $j$th magnon mode.
For a detailed derivation, see \ref{appex:DerivationofHamiltonian}.

\begin{figure}[tb]
    \centering
    \includegraphics[width=1.0\linewidth]{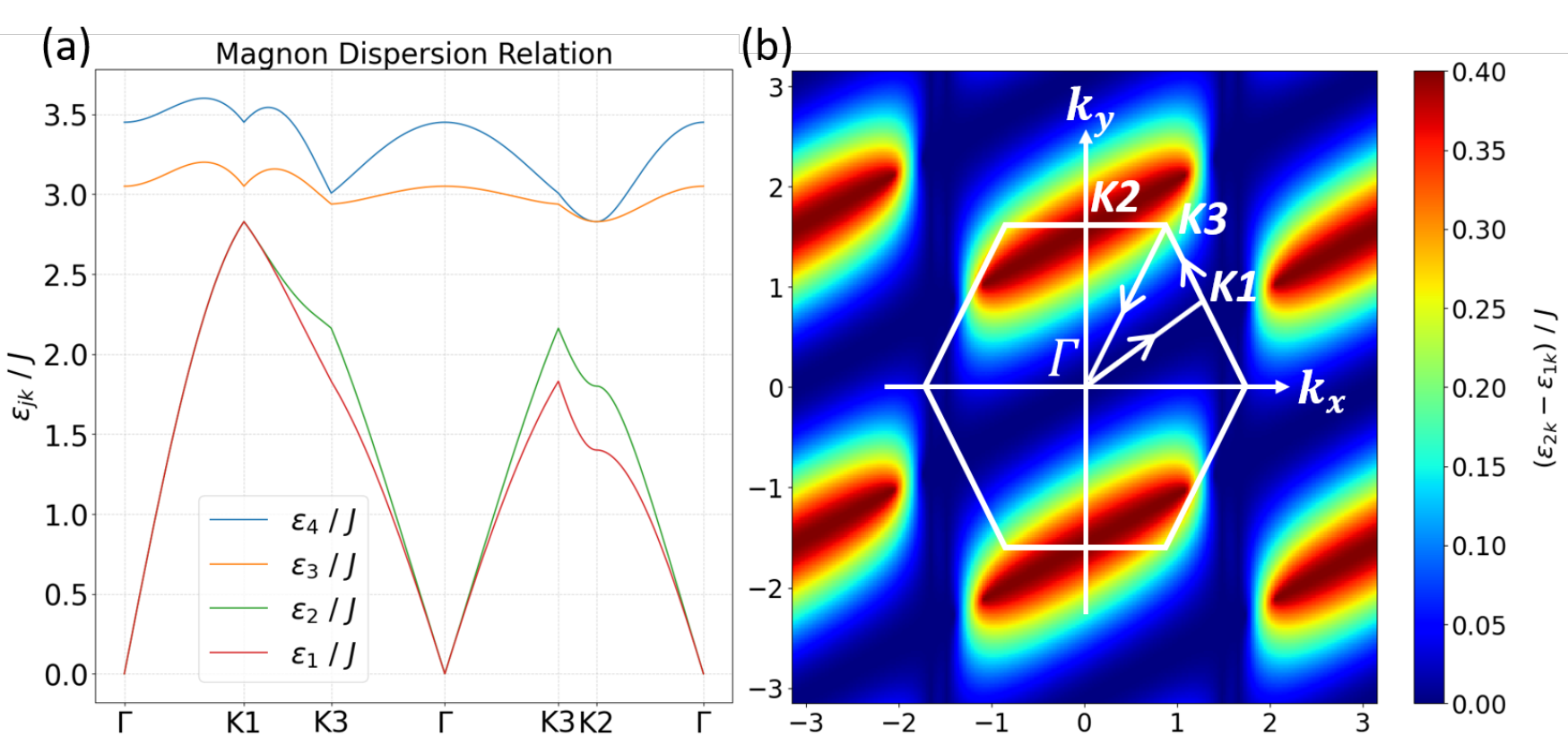}
    \caption{(a) Magnon dispersion relations along the high-symmetry momentum path shown in panel (b). 
    (b) Heatmap of the difference between the lowest and second-lowest magnon bands, $(\varepsilon_2 - \varepsilon_1)/J$ as a function of the wave vector ${\bm k} = (k_x,k_y)$. The white lines denote a representative momentum path in the reduced Brillouin zone (BZ).}
    \label{fig:pathandbanddiagram}
\end{figure}

An example of the magnon 
dispersion is shown in Fig.~\ref{fig:pathandbanddiagram}(a). The parameters are chosen as $J_1 = -0.3J$ and $J_2 = -0.1J$.
We find that two of the four magnon modes have linear dispersion near the $\Gamma$ point, which is a feature common to conventional antiferromagnets. On the other hand, the two lowest magnon modes have slightly different energies.
In Fig.~\ref{fig:pathandbanddiagram}(b), 
we show a contour plot of the difference between the lowest and second-lowest energy bands in the two-dimensional reciprocal space.
In contrast to altermagnets, the energy splitting is isotropic and 
does not exhibit mode crossing as observed in altermagnets.
This isotropic splitting of magnon modes is the main feature of the CF insulator and leads to characteristic spin transport in CF-based junctions.

\subsection{Normal metal and interfacial exchange coupling}

We consider the normal metal and the interfacial exchange  coupling, whose Hamiltonians are given by
\begin{align}
H_{\text{NM}} &= \sum_{{\bm k}\sigma} \xi_{\bm k} c_{{\bm k}\sigma}^\dagger c_{{\bm k}\sigma}, \\
H_{\text{int}} &= \sum_{\bm{r}} J_{\nu {\bm R}, {\bm r}} \bm{S}_{\bm{R}} \cdot \bm{s}_{\bm r}, \label{dotprod}
\end{align}
respectively. Here, $\xi_{\bm k}$ is the electron dispersion measured from the chemical potential, $c_{{\bm k}\sigma}$ is the annihilation operator for conduction electrons with wave vector ${\bm k}$ and spin $\sigma$,
${\bm r}$ denotes a site in the NM neighboring 
the site ${\bm R}$ in the CF, and $\bm{s}_{\bm r}$ is the spin operator of the normal metal at site $\bm{r}$ defined as
\begin{align}
{\bm s}_{\bm r} = \sum_{\bm k,\bm q} \sum_{\sigma,\sigma'} e^{i{\bm q}\cdot {\bm r}} c_{{\bm k + \bm q}\sigma}^\dagger ({\bm \sigma})_{\sigma \sigma'} c_{{\bm k}\sigma'}.
\end{align}
Here, ${\bm \sigma}=(\sigma_x,\sigma_y,\sigma_z)$ denotes the Pauli matrices.

\section{Formulation}
\label{sec:formulation}

The spin current flowing into the NM is defined by
\begin{align}
I_S &= \hbar \dot{s}_{\rm tot}^z = i[s_{\rm tot}^z, H], \\
s_{\rm tot}^z &= \frac12 \sum_{\bm k} (c_{{\bm k}\uparrow}^\dagger c_{{\bm k}\uparrow}-c_{{\bm k}\downarrow}^\dagger c_{{\bm k}\downarrow}).
\end{align}
By substituting the total Hamiltonian into the definition, the spin current can be written as
\begin{align}
I_S &= \frac{i}{2}\sum_{\nu\bm{R}} (J_{\nu{\bm R},{\bm r'}} S_{\nu\bm{R}}^{+} s_{{\bm r}'}^- - {\rm h.c.}).
\end{align}
Within the Keldysh formalism, the interface spin current is evaluated to second order in the interfacial exchange coupling $H_{\rm int}$~\cite{kato19}.
We then perform configurational averaging over 
interface disorder, assuming
\begin{align}
\left\langle J_{\nu{\bm R},{\bm r}} J_{\nu^{\prime}{\bm R}',{\bm r}'}^{\ast} \right\rangle_{\rm imp} = |J|^2 \delta_{\bm{R},\bm{R}^{\prime}} \delta_{\nu,\nu^{\prime}} ,
\label{randave}
\end{align}
where $\langle \cdots \rangle_{\rm imp}$ denotes disorder averaging over interfacial atomic configurations and $J$ is the interfacial exchange amplitude.
The resulting current is expressed in terms of retarded and lesser components of the response functions in CF and NM.
For details of these response functions, see \ref{app:retarded} and \ref{app:lesser}.
Finally, the spin current is given by
\begin{align}
&\langle I_S \rangle=\frac{4\hbar N_{\rm int}|J|^2}{N_{\rm NM} N_{\text{CF}}}  \sum_{\bm{k} ,\bm{q}}  \sum_{\nu} \int_{-\infty}^{\infty} \frac{d\omega}{2\pi}\,\text{Im}\,G_{\nu \nu}^{R}(\bm{k},\omega) \notag \\
&\qquad\times\,\text{Im}\,\chi^{R}(\bm{q},\omega)\bigg(f^{\rm CF}_{\nu\nu}(\hbar \omega) - f^{\rm NM}(\hbar \omega)\bigg),
\label{spincurrent}
\end{align}
where $N_{\rm int}$ is the number of 
exchange bonds at the interface (see~
\ref{appex:derivationofSpinCurrent}).
Here, we define the nonequilibrium distribution functions, $f^{\rm CF}_{\nu\nu}(\hbar \omega)$ and $f^{\rm NM}(\hbar \omega)$, as
\begin{align}
f^{\rm CF}_{\nu\nu}(\hbar \omega) &= \frac{G^<_{\nu\nu}({\bm k},\omega)}{2i\,{\rm Im}\,G_{\nu\nu}^R({\bm k},\omega)} ,\\
f^{\rm NM}(\hbar \omega) &= \frac{\chi^<({\bm k},\omega)}{2i\,{\rm Im}\,\chi^R({\bm k},\omega)} ,
\end{align}
respectively.
Assuming thermal equilibrium in the CF and NM layers, the distribution functions $f^{\rm CF}_{\nu\nu}(\hbar \omega)$ and $f^{\rm NM}(\hbar \omega)$ reduce to the Bose--Einstein distribution functions with temperatures $T_{\rm CF}$ and $T_{\rm NM}$, respectively, according to the fluctuation–dissipation theorem.

We evaluate thermally driven spin current across the CF/NM interface by setting 
$T_{\rm CF}=T+\Delta T$ and $T_{\rm NM}=T$.
To linear order in $\Delta T$, the difference between the distribution functions becomes
\begin{align}
\label{BoseLinearTemperature}
f_{\nu \nu}^{\text{CF}}(\hbar \omega,T + \Delta T) - f^{\text{NM}}(\hbar \omega,T) 
\approx \frac{k_B\beta^2 \hbar \omega}{4\sinh^2(\beta \hbar \omega /2)} \Delta T,
\end{align}
where $\beta = (k_{\rm B}T)^{-1}$.
Substituting this expression into Eq.~(\ref{spincurrent}), we obtain
\begin{align}
\label{SSEspincurrent}
\langle I_S^{\rm SSE} \rangle
&= \frac{4S_0A k_{\rm B}\Delta T}{N_{\rm CF}}  \sum_{\bm{k}\in \Lambda}\sum_{j =1}^8\sum_{\nu =1}^4 \frac{(\beta\varepsilon_{\bm{k},j})^2}{ \sinh^2(\beta \varepsilon_{\bm{k},j}/2)}  \sigma_\nu \sigma_j|(T_{\bm{k}})_{\nu j}|^2 ,
\end{align}
where $A=4\pi N_{\rm int}|J|^2 N(0)^2$, $\sigma_\nu = 1$ ($-1$) for $\nu=1,2$ ($\nu=3,4$), and $\sigma_j = 1$ ($-1$) for $j=1,2,3,4$ ($j=5,6,7,8$)
(see also \ref{app:retarded}).

To benchmark the CF results, we derive the corresponding SSE spin current for a ferromagnetic insulator (FI)/NM junction by applying the same formalism to a single-sublattice ferromagnet with quadratic magnon dispersion $\lambda_{\bm{k}} = J_{\text{FI}}S_0a^2|\bm{k}|^2$~\cite{Kittel2005}, where $J_{\text{FI}}$ and $a$ denote the exchange coupling and lattice constant of the FI, respectively.
The resulting spin current reads
\begin{align}
\label{ferromagneticspinseebeck}
\langle I_{S,0}^{\rm SSE} \rangle
&= \frac{4S_0A k_{\rm B}\Delta T}{N_{\rm FI}} \sum_{\bm{k}\in \Lambda}\frac{(\beta \lambda_{\bm{k}})^2}{ \sinh^2(\beta \lambda_{\bm{k}}/2)} .
\end{align}

\section{Result}
\label{sec:result}

\begin{figure}[tb]
    \centering
    \includegraphics[width=1.0\linewidth]{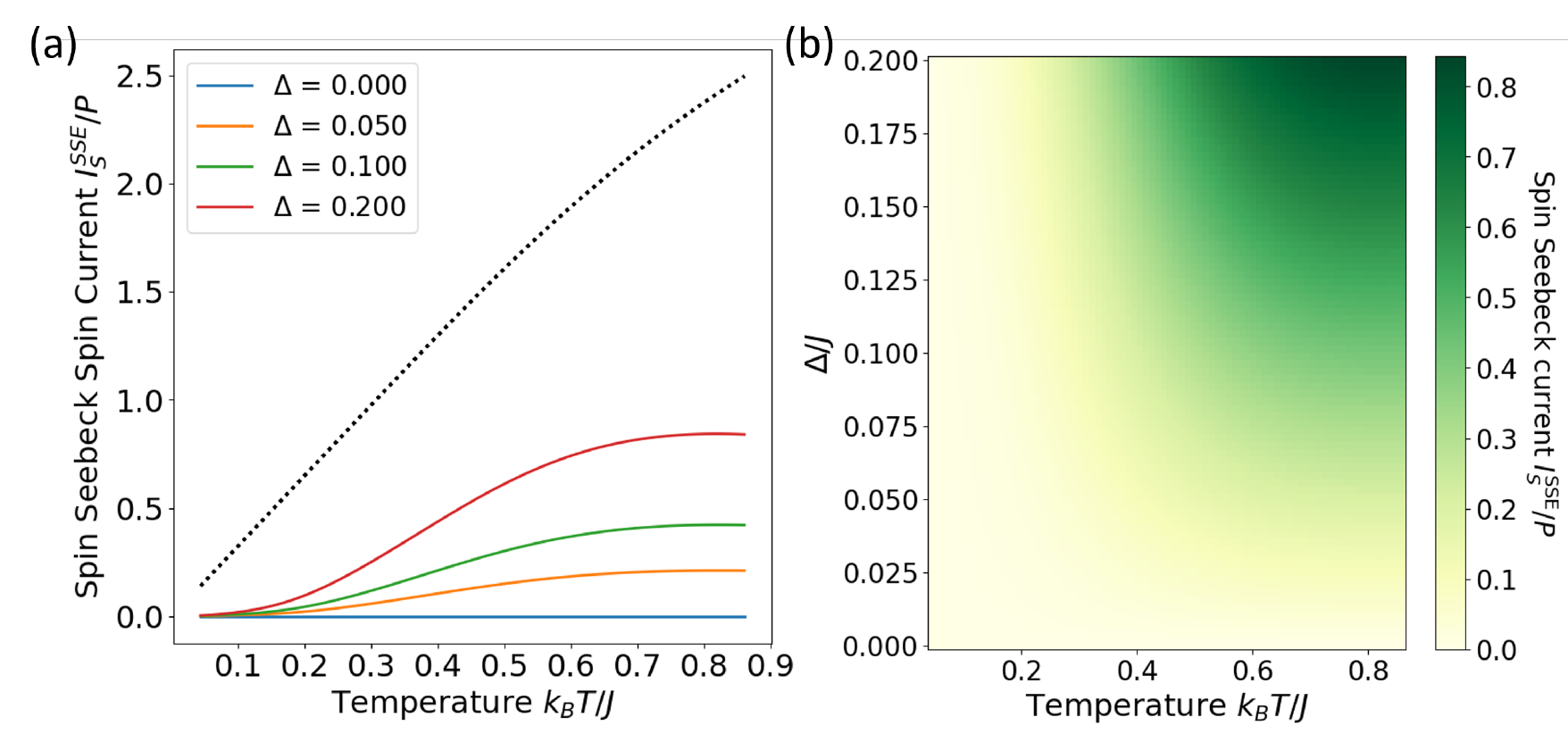}
    \caption{(a) Spin current induced by a temperature bias, plotted as a function of temperature.   
    Solid lines show the spin current in the compensated ferrimagnet/normal-metal junction, while the dotted line represents that in the FI/normal-metal junction.
    The exchange interactions are chosen to be $J_1 = -0.5J + \Delta$ and $J_2 = -0.5J - \Delta$, where $\Delta$ takes the values $0.0J$, $0.05J$, $0.1J$, and $0.2J$. Both the solid and dotted lines are normalized by $ P  = 4S_0A k_{\rm B}\Delta T$. (b) Heatmap of the spin Seebeck current $I_S^{\rm SSE}/P$ as a function of temperature $k_{\rm B}T/J$ and exchange difference $\Delta /J$.}
    \label{fig:spincurrent}
\end{figure}

The solid curves in Fig.~\ref{fig:spincurrent}(a) show the thermally induced spin current in CF/NM for the same exchange settings as in Fig.~\ref{fig:pathandbanddiagram}.
At low temperature, the current follows an approximate $T^2$ dependence, then reaches a maximum when thermal occupation extends over the magnon bandwidth 
set by $J$.
The dotted FI/NM curve serves as a benchmark.
Over a broad range, the signal for CF/NM remains comparable to that of FI/NM in magnitude.
This is in stark contrast to the anisotropy-based model studied in Ref.~\cite{Lee2025}, where the spin Seebeck signal is approximately two orders of magnitude smaller than the ferromagnetic reference.
The key difference lies in the energy scale of the magnon splitting: in the anisotropy-based model, the splitting is set by the single-ion anisotropy $\Delta K$, which is typically much smaller than the exchange coupling $J$.
In the present exchange-coupling-based model, the splitting is governed by $|J_1-J_2|$, which can be of the same order as $J$. 
This larger splitting leads to a greater thermal population imbalance between the two magnon branches, thereby producing a significantly enhanced spin current.
In Fig.~\ref{fig:spincurrent}(b), we also show a contour plot of the spin current as a function of temperature $T$ and the exchange difference  $\Delta = (J_1 - J_2)/2$, while keeping the average $(J_1+J_2)/2$ fixed at $-J/2$.
We find that the peak around $k_{\rm B}T/J = 0.8$ increases as the exchange difference $\Delta$ increases.

\begin{figure}[tb]
    \centering
    \includegraphics[width=0.5\linewidth]{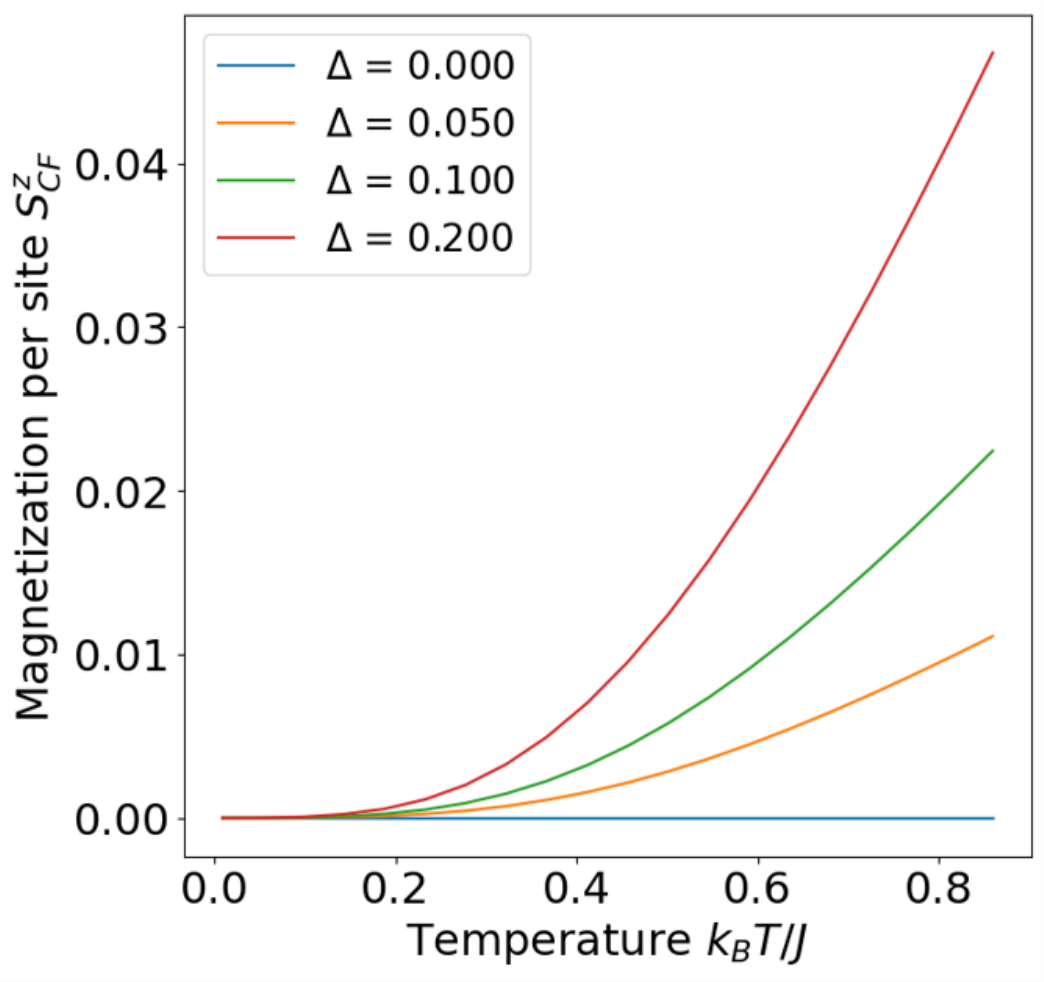}
    \caption{Temperature dependence of the magnetization per unit cell, $S^z_{\rm CF} = \sum_{\nu=1}^4 \langle S_{\nu}^z  \rangle$ for different values of the exchange difference $\Delta/J$. 
    }
    \label{fig:magnetization}
\end{figure}

We note that although the net magnetization of the CF vanishes at zero temperature by construction, a finite magnetization $M_{\rm CF}(T)$ is induced at finite temperatures due to the asymmetric thermal population of the split magnon branches, as shown in Fig.~\ref{fig:magnetization}.
The induced magnetization increases with increasing $\Delta = (J_1-J_2)/2$ and temperature, reflecting the enhanced imbalance of magnon occupation.
This thermally induced magnetization is a characteristic feature of the exchange-coupling-based CF model and is substantially larger than that in the anisotropy-based model of Ref.~\cite{Lee2025}, where it was estimated to be on the order of $10^{-2}$ per unit cell.
However, the magnitude of this magnetization is too small to account for the spin Seebeck effect within the conventional picture based on finite magnetization.
This indicates that the spin Seebeck effect in the CF/NM junction is intrinsic to the compensated ferrimagnet.

Finally, we briefly discuss the origin of the spin current in the CF/NM junction.
The amplitudes of magnon modes depend on the sublattice and are represented by $|T_{\nu j}|$, where $j$ is the magnon mode index and $\nu$ denotes the sublattice.
For simplicity, we focus on the two lowest magnon modes, whose energies are given by $\varepsilon_1(\bm{q})$ and $\varepsilon_2(\bm{q})$.
The corresponding wave functions have imbalanced weights across the sublattices (i.e., $|T_{\nu j}|\ne |T_{\nu' j}|$ for $\nu\ne \nu'$), leading to a finite spin current even when the interfacial exchange coupling is identical for all the sublattices.
We recover the vanishing spin current for 
$J_1 = J_2$ (see the blue solid line in Fig.~\ref{fig:spincurrent}(a)), since the system effectively reduces to a conventional antiferromagnet~\cite{Ohnuma2013}.

It is instructive to compare the spin Seebeck effect in CF/NM junctions with that in altermagnet (AM)/NM junctions.
Using Eq.~\eqref{SSEspincurrent} and the Bogoliubov transformation with a paraunitary matrix for a two-sublattice altermagnet~\cite{Cui2023},
\begin{align}
    T_{\bm{k}} =
    \begin{pmatrix}
        u_{\bm{k}} & v_{\bm{k}}^{\ast}\\
        v_{\bm{k}} & u_{\bm{k}}^{\ast}
    \end{pmatrix},
\end{align}
the spin current can be written as
\begin{align}
    I_{S,{\rm ALM}}^{\rm SSE} &\propto \sum_{\bm{k} \in \Lambda} \bigg[\frac{\varepsilon_{\bm{k},A}^2}{ \sinh^2(\beta \varepsilon_{\bm{k},A}/2)} - \frac{\varepsilon_{\bm{k},B}^2}{ \sinh^2(\beta \varepsilon_{\bm{k},B}/2)}\bigg]\bigg(|u_{\bm{k}}|^2 + |v_{\bm{k}}|^2\bigg).
\end{align}
Because the magnon splitting in altermagnets is anisotropic ($d$-wave-like), the two branches satisfy $\varepsilon_{\bm{k},A}=\varepsilon_{\bm{k},B}$ along nodal lines and exchange their roles under rotation.
As a result, the momentum summation over the 
Brillouin zone yields a vanishing spin current, $I_{S,{\rm AM}}^{\rm SSE}=0$.
This stands in sharp contrast to the CF case, where the isotropic ($s$-wave-like) splitting ensures $\varepsilon_1(\bm{k})\ne \varepsilon_2(\bm{k})$ throughout the Brillouin zone (except at the $\Gamma$ point), resulting in a finite spin current after the momentum summation.
Thus, among magnetically compensated systems, compensated ferrimagnets are uniquely suited for thermal spin-current generation at interfaces with uniform exchange coupling.

\section{Discussion and Summary}
\label{sec:summary}

We have investigated the spin Seebeck effect in compensated ferrimagnet/normal-metal junctions.
Our analysis shows that sublattice inequivalence controls isotropic magnon splitting and enables efficient interfacial spin-current generation despite zero net magnetization.
The present work employs a four-sublattice exchange-coupling-based model, in which sublattice inequivalence is realized through inequivalent ferromagnetic exchange couplings $J_1\ne J_2$
rather than through unequal single-ion anisotropies, as in our previous study~\cite{Lee2025}.
This distinction has important physical consequences: the magnon splitting is set by $|J_1-J_2|$ rather than 
$\Delta K$, enabling a splitting on the scale of the exchange coupling itself.
As a result, the spin Seebeck signal in the present model reaches the same order of magnitude as that in ferromagnetic junctions, in contrast to the anisotropy-based model, where it is approximately two orders of magnitude smaller.

We have also shown that the spin Seebeck effect vanishes identically in altermagnet/normal-metal junctions with uniform interfacial exchange coupling, owing to the anisotropic ($d$-wave-like) character of the magnon splitting.
This highlights the unique advantage of compensated ferrimagnets, whose isotropic ($s$-wave-like) splitting ensures a finite spin current after integration over the full Brillouin zone.
Combined with the vanishing spin current in conventional antiferromagnets ($J_1=J_2$), these results establish a clear hierarchy among magnetically compensated systems for interfacial thermal spin transport: compensated ferrimagnets generate a finite and potentially large spin Seebeck signal, whereas both altermagnets and conventional antiferromagnets yield vanishing signals under the same conditions.

The exchange-coupling-based mechanism studied here is particularly relevant for organic antiferromagnets~\cite{Kawamura_PRL2024} and two-dimensional van der Waals heterostructures~\cite{Liu_PRL2025}, where inequivalent molecular or layer environments naturally give rise to different exchange pathways.
Since exchange couplings in these systems are more readily tunable than single-ion anisotropies, the present model provides 
practical guidelines for designing materials that maximize the spin Seebeck response.

\section*{Acknowledgments}

The authors are grateful to R. Sano for valuable discussions.
This work was supported by the National Natural Science Foundation of China (NSFC) under Grant No. 12374126, 
by the Priority Program of Chinese Academy of Sciences under Grant No. XDB28000000, and by JSPS KAKENHI Grant Nos. JP20H00122, JP21H01800, JP21H04565, JP23H01839, JP23H03818, JP24K06951, and JP24H00322 from MEXT, Japan. TM is supported by JST FOREST Grant No. JPMJFR236N.

\appendix

\section{Spin-Wave Approximation}
\label{appex:DerivationofHamiltonian}

This appendix provides a brief derivation of the matrix form in Eq.~(\ref{modelequation}) from the original Hamiltonian given in Eq.~(\ref{rawmodelequation}).
By employing the spin-wave approximation given in Eqs.~(\ref{spinwaveapprox1})-(\ref{spinwaveapprox3}) and 
using the Fourier transform 
\begin{align}
    b_{\nu \bm{R}} = \frac{1}{\sqrt{N_{\text{CF}}}} \sum_{\bm{k} \in \Lambda} e^{i(\bm{k}\cdot(\bm{R} + \bm{\delta_\nu}))} b_{\nu \bm{k}},
\end{align}
the exchange term in the Hamiltonian, Eq.~(\ref{rawmodelequation}), can be rewritten as
\begin{align}
&\sum_{\nu,\bm{R}}\sum_{\mu,\bm{R}^{\prime}}J_{\bm{R},\bm{R}^{\prime}}^{\nu \mu} \bm{S}_{\nu \bm{R}} \cdot \bm{S}_{\mu \bm{R}^{\prime}} \notag \\
    &= S_0\sum_{\bm{k} \in \Lambda} \bigg[3J(b_{1 \bm{k}}^{\dagger}b_{1 \bm{k}} + b_{2 \bm{k}}^{\dagger}b_{2 \bm{k}} + b_{3 \bm{k}}^{\dagger}b_{3 \bm{k}} + b_{4 \bm{k}}^{\dagger}b_{4 \bm{k}}) \notag \\
    & \qquad - J_2(b_{3 \bm{k}}^{\dagger}b_{3 \bm{k}} + b_{4 \bm{k}}^{\dagger}b_{4 \bm{k}}) - J_1(b_{1 \bm{k}}^{\dagger}b_{1 \bm{k}} + b_{2 \bm{k}}^{\dagger}b_{2 \bm{k}}) \notag \\
    & \quad + J\bigg(e^{i\bm{k} \cdot \bm{\delta_x}} b_{4 \bm{k}}^{\dagger}b_{2,- \bm{k}}^{\dagger} + e^{-i\bm{k} \cdot \bm{\delta_x}} b_{4 \bm{k}}b_{2,- \bm{k}} + e^{i\bm{k} \cdot \bm{\delta_x}} b_{3 \bm{k}}^{\dagger}b_{1,- \bm{k}}^{\dagger} + e^{-i\bm{k} \cdot \bm{\delta_x}} b_{3 \bm{k}}b_{1,- \bm{k}} \notag \\
    & \qquad + e^{-i\bm{k} \cdot \bm{\delta_x}} b_{1 \bm{k}}b_{3,- \bm{k}} + e^{i\bm{k} \cdot \bm{\delta_x}} b_{1 \bm{k}}^{\dagger}b_{3,- \bm{k}}^{\dagger} +e^{-i\bm{k} \cdot \bm{\delta_x}} b_{2 \bm{k}}b_{4,- \bm{k}} + e^{i\bm{k} \cdot \bm{\delta_x}} b_{2 \bm{k}}^{\dagger}b_{4,- \bm{k}}^{\dagger} \notag \\
    & \qquad + e^{-i\bm{k} \cdot \bm{\delta_y}} b_{2 \bm{k}}b_{3,- \bm{k}} + e^{i\bm{k} \cdot \bm{\delta_y}} b_{2 \bm{k}}^{\dagger}b_{3,- \bm{k}}^{\dagger} + e^{i\bm{k} \cdot \bm{\delta_y}} b_{4 \bm{k}}^{\dagger}b_{1,- \bm{k}}^{\dagger} + e^{-i\bm{k} \cdot \bm{\delta_y}} b_{4 \bm{k}}b_{1,- \bm{k}}\bigg) \notag \\
    & \quad + J_2(e^{i\bm{k} \cdot \bm{\delta_y}} b_{3 \bm{k}}^{\dagger}b_{4 \bm{k}} + e^{-i\bm{k} \cdot \bm{\delta_y}} b_{3\bm{k}}b_{4\bm{k}}^{\dagger}) + J_1(e^{i\bm{k} \cdot \bm{\delta_y}} b_{2 \bm{k}}^{\dagger}b_{1 \bm{k}} + e^{-i\bm{k} \cdot \bm{\delta_y}} b_{2\bm{k}}b_{1\bm{k}}^{\dagger})\bigg] .
\end{align}
By introducing the basis 
\begin{align}
    {\bm \Psi}=(b_{1{\bm k}},b_{2{\bm k}},b_{3{\bm k}},b_{4{\bm k}},b_{1-{\bm k}}^\dagger,b_{2-{\bm k}}^\dagger,b_{3-{\bm k}}^\dagger,b_{4-{\bm k}}^\dagger)^t,
\end{align} 
the Hamiltonian of the CF is rewritten as
\begin{align}
H_{\rm CF} &= \sum_{\bm k} {\bm \Psi}^\dagger M_{\bm{k}} {\bm \Psi}, \\
M_{\bm{k}} &= 
\begin{pmatrix}
A_{\bm k} & B_{\bm k} \\
B_{\bm k}^\ast & A_{\bm k} 
\end{pmatrix} , \\
A_{\bm k} &= 
\begin{pmatrix}
0 & 0 & 2J\cos k_x a & Je^{- i k_y a}\\
0 & 0 & Je^{ i k_y a} & 2J\cos k_x a \\ 
2J\cos k_x a & Je^{- i k_y a} & 0 & 0\\
Je^{i k_y a} & 2J\cos k_xa & 0 & 0
\end{pmatrix} , \\
B_{\bm k} &=
\begin{pmatrix}
3J - J_1 & J_1e^{i k_y a} & 0 & 0 \\
J_1e^{-i k_y a} & 3J - J_1 & 0 & 0 \\
0 & 0 & 3J - J_2 & J_2e^{i k_y a} \\
0 & 0 & J_2e^{-i k_y a} & 3J - J_2
\end{pmatrix} .
\end{align}

\section{Retarded response functions}
\label{app:retarded}

The retarded spin susceptibility of the NM is defined as
\begin{align}
    \chi^{R}(\bm{q},t) = \frac{i\theta(t)}{\hbar N_{\rm NM}} \langle [s_{\bm{q}}^+(t),s_{\bm{q}}^{-}(0)]\rangle,
\end{align}
where $N_{\rm NM}$ is the number of unit cells. Similarly, the retarded spin susceptibility in the CF for sublattice $\nu$ is defined as
\begin{align}
\label{definitionofRetarded}
    G^{R}_{\nu\nu}(\bm{k},t) = -\frac{i}{\hbar}\langle [S_{\nu \bm{k}}^+(t),S_{\nu \bm{k}}^{-}(0)] \rangle .
\end{align}
We note that this response function corresponds to the magnon Green's function. Employing the linear spin-wave approximation based on the Holstein-Primakoff transformation, we obtain the magnon retarded Green's function 
\begin{align}
\label{retardedgreenfunction2}
G_{\nu\nu}^{R}(\bm{k},\omega) = 
\left\{ \begin{array}{ll}
\displaystyle{\sum_{j = 1}^8 \frac{2S_0 \sigma_j|(T_{\bm{k}})_{\nu j}|^2}{\hbar \omega - \sigma_j \varepsilon_{\bm{k},j} + i \delta},} & (\nu = 1,2), \\
\displaystyle{-\sum_{j = 1}^8 \frac{2S_0 \sigma_j|(T_{\bm{k}})_{\nu j}|^2}{\hbar \omega + \sigma_j \varepsilon_{\bm{k},j} + i \delta},} & (\nu = 3,4), \end{array} \right.
\end{align}
where $\sigma_j = \pm 1$ corresponds to the particle or hole branch in the Nambu basis and $\delta$ is a positive infinitesimal. 
For further details, see \ref{appex:DerivationofGreenFunction}.
To incorporate Gilbert damping phenomenologically, $\delta$ is replaced by $\alpha \hbar \omega$, where $\alpha$ is the Gilbert damping constant.

\section{Lesser response functions}
\label{app:lesser}

We define the lesser components for the CF and NM as
\begin{align}
\label{chilesser}
\chi^{<}(\bm{q},t) &= -\frac{i}{\hbar N_{\rm NM}} \langle s_{\bm{q}}^{-}(0)s_{\bm{q}}^+(t) \rangle,\\
\label{Glesser}    
G^{<}_{\nu\nu}(\bm{k},t) &= \frac{i}{\hbar}\langle S_{\nu \bm{k}}^{-}(0) S_{\nu \bm{k}}^+(t) \rangle .
\end{align}
When the CF and NM are decoupled and 
in thermal equilibrium at temperatures $T_{\rm CF}$ and $T_{\rm NM}$, the lesser components can be related to the retarded components through the fluctuation-dissipation theorem as~\cite{Stefanucci2013}\red{:}
\begin{align}
\label{KMSlesserRetarded}
G^{<}_{\nu \nu}(\bm{k},\omega) &=2if(\hbar \omega,T_{\rm CF}) \text{Im}\, G^{R}_{\nu\nu}(\bm{k},\omega), \\
\chi^{<}(\bm{q},\omega) &= 2if(\hbar \omega,T_{\rm NM}) \text{Im}\, \chi^{R}(\bm{q},\omega),
\label{KMSlesserRetarded2}
\end{align} 
where $f(\epsilon,T)=(e^{\epsilon/k_{\rm B}T}-1)^{-1}$ is the Bose--Einstein distribution function.
These expressions are used for formulating the spin current induced by the SSE in Sec.~\ref{sec:formulation}.

\section{Detailed Derivation of the Spin Current}\label{appex:derivationofSpinCurrent}

We derive Eq.~(\ref{spincurrent}) from the contour-ordered Keldysh expression~\cite{Bruus,kato19}.
\begin{align}
\langle I_S(t) \rangle 
&= \text{Im}\,\bigg[ \sum_{\nu \bm{R}}J_{\nu {\bm R},{\bm r}} \bigg\langle S_{\nu \bm{R}}^+(\tau_1)s_{\bm r}^-(\tau_2) \exp\bigg(-\frac{i}{\hbar}\int_C d\tau H_{\text{int}}(\tau) \bigg)\bigg\rangle_0\bigg] \notag \\
&= -\frac{2}{\hbar} \, \text{Re}\,\bigg[ \sum_{\nu \bm{R}} \sum_{\nu^{\prime} \bm{R}^{\prime}} J_{\nu {\bm R},{\bm r}} J_{\nu^{\prime} {\bm R}^{\prime}, {\bm r}'}^{\ast} \int_C d\tau \, \bigg\langle \mathcal{T} S_{\nu \bm{R}}^+(\tau_1) s_{{\bm r}}^-(\tau_2) S_{\nu^{\prime} \bm{R}^{\prime}}^- s_{{\bm r}'}^+(\tau)\bigg\rangle_0\bigg] ,
\end{align}
where $A(\tau) = e^{\tau H_0/\hbar} A e^{-\tau H_0/\hbar}$ denotes the contour-time evolution of operator $A$ under $H_0 = H_{\rm CF} + H_{\rm NM}$, $\langle \cdots \rangle_{0}$ is the thermal average with respect to $H_0$, and $\tau_1=(t,+)$, $\tau_2=(t,-)$ are contour times.
Using Eq.~\eqref{randave}, we obtain
\begin{align}
\langle I_S\rangle &= -2\hbar \,\text{Re}\,\bigg[ \sum_{\nu \bm{R}} \sum_{\bm{k} ,\bm{q}} \frac{|J|^2}{N_{\rm CF}N_{\rm NM}} \int_C d\tau \, G_{\nu \nu} (\bm{q},\tau_1,\tau) \chi(\bm{k},\tau,\tau_2)\bigg] \notag \\
&= -2\hbar\sum_{\bm{k} ,\bm{q}}  \sum_{\nu}\,\text{Re}\, \bigg[  \sum_{\nu \bm{R}}  \frac{|J|^2}{N_{\rm CF}N_{\rm NM}} \notag \\
&\qquad \qquad \times \int_{-\infty}^{\infty} \frac{d\omega}{2\pi} \, \bigg(\chi^{<}(\bm{k},\omega)G_{\nu \nu}^{R,(a)}(\bm{q},\omega) + \chi^{A}(\bm{k},\omega) G_{\nu \nu}^{<,(a)}(\bm{q},\omega)\bigg)\bigg],
\end{align}
where the second line follows from the Langreth rules~\cite{Stefanucci2013}.
Finally, substituting Eqs.~\eqref{KMSlesserRetarded} and \eqref{KMSlesserRetarded2} yields Eq.~(\ref{spincurrent}).

\section{Magnon Green's Functions}\label{appex:DerivationofGreenFunction}

In this appendix, we derive the explicit form of the retarded magnon Green's function for the CF, $G_{\nu\nu}^R(\bm{k},\omega)$.
We first consider the case 
$\nu = 1,2$.
The retarded Green's function is given by
\begin{align}
\label{retardedgreensfunction}
G_{\nu\nu}^R(\bm{k},t) &= -i \frac{2S_0}{\hbar} \theta(t) \langle[b_{\nu \bm{k}}(t),b_{\nu \bm{k}}^{\dagger}(0)] \rangle \notag \\ 
&= -i \frac{2S_0}{\hbar} \theta(t) \bigg \langle \bigg[ \sum_{j = 1}^8(T_{\bm{k}})_{\nu j }\Phi_j(t),\sum_{i = 1}^8(T_{\bm{k}}^{\ast})_{\nu i}(\Phi^{\dagger 
})_i\bigg]\bigg \rangle \notag \\
&= -i \frac{2S_0}{\hbar} \theta(t) \sum_{j = 1}^8 \sigma_j|(T_{\bm{k}})_{\nu j}|^2e^{-i\sigma_j\varepsilon_{\bm{k},j}t/\hbar} ,
\end{align}
where $\Phi_j(t) = e^{iHt/\hbar} \Phi_j e^{-iHt/\hbar} = e^{-i\sigma_j\varepsilon_{\bm{k},j}t/\hbar} \Phi_j$ has been used.
By the Fourier transformation, we obtain
\begin{align}
  G_{\nu\nu}^R(\bm{k},\omega) &= 2S_0 \sum_{j = 1}^8 \frac{\sigma_j|(T_{\bm{k}})_{\nu j}|^2}{\hbar \omega - \sigma_j \varepsilon_{\bm{k},j} + i \alpha \hbar \omega},
\end{align}  
where $\delta$ is replaced by $\alpha \hbar\omega$ to take into account the Gilbert damping phenomenologically.
The result for $\nu = 3,4$ can be obtained in a similar manner.

For completeness, we also derive the retarded Green's function of the FI, defined as
\begin{align}
    G^R(\bm{k},t) &= -i\frac{2S_0}{\hbar} \theta(t) \langle [ b_{\bm k}(t),b_{\bm k}^\dagger(0)] \rangle.
\end{align}
We note that the Hamiltonian of the FI is written in quadratic form as $H_{\rm FI} = \sum_{{\bm k}} \varepsilon_{\bm k} b_{\bm k}^\dagger b_{\bm k}$ with $\varepsilon_{\bm{k}} = JS_0a^2 |\bm{k}|^2$, where $J$ is the exchange interaction in the bulk FI, $S_0$ is the magnitude of the localized spin, and $a$ is the lattice spacing.
Then, the retarded Green's function is calculated as
\begin{align}
G^R(\bm{k},\omega) &= \int dt \,  G^R(\bm{k},t) e^{i\omega t} = \frac{2S_0}{\hbar \omega - \varepsilon_{\bm{k}} + i \alpha \hbar \omega} ,
\end{align}
where the positive infinitesimal $\delta$ is replaced by $\alpha \hbar \omega$.
Substituting $\mathrm{Im}\, G^R(\bm{k},\omega)$ from the above expression together with Eq.~\eqref{BoseLinearTemperature} into Eq.~\eqref{spincurrent}, we obtain Eq.~\eqref{ferromagneticspinseebeck}.

\bibliographystyle{elsarticle-num} 
\bibliography{reference}
\end{document}